\documentclass[aps,prb,showpacs,floatfix,superscriptaddress,twocolumn]{revtex4-1}
\usepackage{graphicx}    
\usepackage{hyperref}    
\usepackage{bm}          	
\usepackage[sumlimits,intlimits]{amsmath}
\usepackage{amsfonts,amssymb}
\usepackage[section]{placeins}

\begin{document}

\title{Collective modes in anisotropic double layer systems}

\author{A. S. Rodin}
\affiliation{Centre for Advanced 2D Materials and Graphene Research Centre,  National University of Singapore, 6 Science Drive 2, 117546, Singapore}
\author{A. H. Castro Neto}

\affiliation{Centre for Advanced 2D Materials and Graphene Research Centre,
 National University of Singapore, 6 Science Drive 2, 117546, Singapore}
\affiliation{Boston University, 590 Commonwealth Ave., Boston MA 02215}

\date{\today}
\begin{abstract}
We study collective modes in anisotropic double layer systems. To this end, we derive a zero-$T$ dynamic polarization function for a conductor with a parabolloidal dispersion. Following this, we demonstrate the dependence of the plasmonic modes on the relative orientation, doping, and system anisotropy.

\end{abstract}


\maketitle

\section{Introduction}
The field of two-dimensional (2D) heterostructures is currently experiencing a rapid development owing to the improved manufacturing and manipulation techniques.~\cite{Novoselov2012tdc,Geim2013vdw} Most commonly used ingredients in these composite systems include graphene, boron nitride, and transition metal dichalcogenides. A recent addition to the catalogue of 2D materials is monolayer black phosphorus, also referred to as phosphorene.~\cite{Xia2014rbp,Lu2014paf,Castellanos_Gomez2014iac,Koenig2014eff}

Phosphorene is a semiconductor with an almost direct gap~\cite{Rodin2014sig,Li2014eah} and a highly anisotropic dispersion.~\cite{Liu2014pau,Li2014bpf,Rodin2014sig,Castellanos_Gomez2014iac,Peng2014sed,Fei2014set,Li2014eah} The band structure of this novel material is highly sensitive to strain and deformation~\cite{Liu2014pau,Rodin2014sig,Peng2014sed,Fei2014set}, making it a good potential candidate for electro-mechanical applications. A recent work~\cite{Low2014pas} addressed the many-body properties of a doped monolayer black phosphorus by analyzing its polarization function. From this, the authors demonstrated that the plasmonic branch in phosphorene is highly anisotropic. In this work, we tackle the problem of collective excitations in coupled rotationally misaligned systemes with parabolloidal bands. We develop a general formalism applicable for any number of layers, but we focus primarily on two-layer configurations. While black phosphorus represents a good example of a system to which our analysis applies, our results are given in the most general form to make them easily adaptable to other materials.

\section{Polarization Function}
The system in question consists of $N$ 2D layers separated by distance $d$. In order to prevent the interlayer charge transfer, dielectric spacers are used to isolate the stack components from each other. Each layer in the stack can have its own dispersion, chemical potential, and orientation. 

To determine the collective modes in a system, one needs to obtain the zeros of the dielectric function. For a multilayer setup, this function is given by a generalized dielectric tensor $\tensor{\varepsilon}$~\cite{Hwang2009pmo,Zhu2013pei} with
\begin{equation}
\varepsilon_{nl}(k,\omega) = \delta_{nl}-V_{nl}(k)\Pi_n(k,\omega)\,,
\label{eqn:Dielectric_Tensor}
\end{equation}
where $\Pi_n(k,\omega)$ is the polarization function of the $n$th layer and $V_{nl}(k)$ is the coupling between $n$th and $l$th layers. Setting the determinant of $\tensor\varepsilon$ equal to zero allows one to obtain the mode dispersion.

In Eq.~\eqref{eqn:Dielectric_Tensor}, the interaction term is given by
\begin{equation}
V_{nl}(k) =\frac{2\pi e^2}{k}\left(e^{-d k}\right)^{|n-l|}
\label{eqn:Interaction}
\end{equation}
and the polarization function is the usual RPA bubble
\begin{equation}
\Pi_n(k,\omega)= \frac{2}{L^D}\sum_\mathbf{q}\frac{n_F(\xi_{\mathbf{q}+\mathbf{k}})-n_F(\xi_{\mathbf{q}})}{\xi^n_{\mathbf{q}+\mathbf{k}}-\xi^n_{\mathbf{q}}-\hbar\omega-i\hbar\eta}\,.
\label{eqn:Pi_General}
\end{equation}
The parameter $\eta$ introduces broadening to the system. Since we are looking for low-energy modes, Eq.~\eqref{eqn:Pi_General} includes only one band as inter-band transitions require energies which are too large. Therefore, we have
\begin{align}
 \xi^n_\mathbf{q}& = \frac{\hbar^2}{2}\left(\frac{q_x^2}{m_x}+\frac{q_y^2}{m_y}\right)-\mu_n\,,
\label{eqn:Dispersion}
\end{align}
where $m_{x/y}$ is the direction-dependent mass and $\mu_n$ is the chemical potential of the layer.

The first step is computing the polarization function $\Pi_n(k,\omega)$. Previously, a static polarization function was calculated at $T = 0$.~\cite{Low2014pas} Here, we determine the finite--$\omega$  result. It is convenient to shift the variables to get
\begin{align}
\Pi_n(k,\omega)& = \frac{2}{L^D}\sum_\mathbf{q}n_F(\xi_{\mathbf{q}})\left[\frac{1}{\xi^n_{\mathbf{q}}-\xi^n_{\mathbf{q}-\mathbf{k}}-\hbar\hat\omega}\right.
\nonumber
\\
&\left.-\frac{1}{\xi^n_{\mathbf{q}+\mathbf{k}}-\xi^n_{\mathbf{q}}-\hbar\hat\omega}\right]\,,
\label{eqn:Pi_Shift}
\end{align}
where $\hat\omega = \omega+i\eta$.

Next, to get rid of the anisotropy, we introduce a change of variables
\begin{equation}
p_{x/y} = \frac{\hbar q_{x/y}}{\sqrt{2m_{x/y}}}\,,\quad s_{x/y} = \frac{\hbar k_{x/y}}{\sqrt{2m_{x/y}}}\,.
\label{eqn:Momentum_Var_Change}
\end{equation}

As was stated earlier, we are interested in rotationally misaligned layers. This means that we need a way to denote momenta in every layer using a common coordinate system. To this end, each layer is assigned the rotational angle $\tau_n$ which is the angle between the $x$-axis in the lab frame and the $x$-axis of the $n$th layer. Denoting $\mathbf{k} = k(\cos\theta,\,\sin\theta)$, where $\theta$ that the vector $\mathbf k$ makes with the $x$-axis in the lab frame, allows us to write
\begin{equation}
s = k\underbrace{\frac{\hbar}{\sqrt{2m_x}}}_{\sqrt{Z}}\underbrace{\sqrt{\cos^2(\tau_n-\theta)+M\sin^2(\tau_n-\theta)}}_{\sqrt{f_n}}\,,
\label{eqn:s}
\end{equation}
where $M = m_x/m_y$. In Eq.~\eqref{eqn:s}, the parameter $Z$ sets up the energy scale of the problem and the orientation factor $f_n$ captures the system anisotropy.

\begin{figure}[h]
\includegraphics[width = 3.5in]{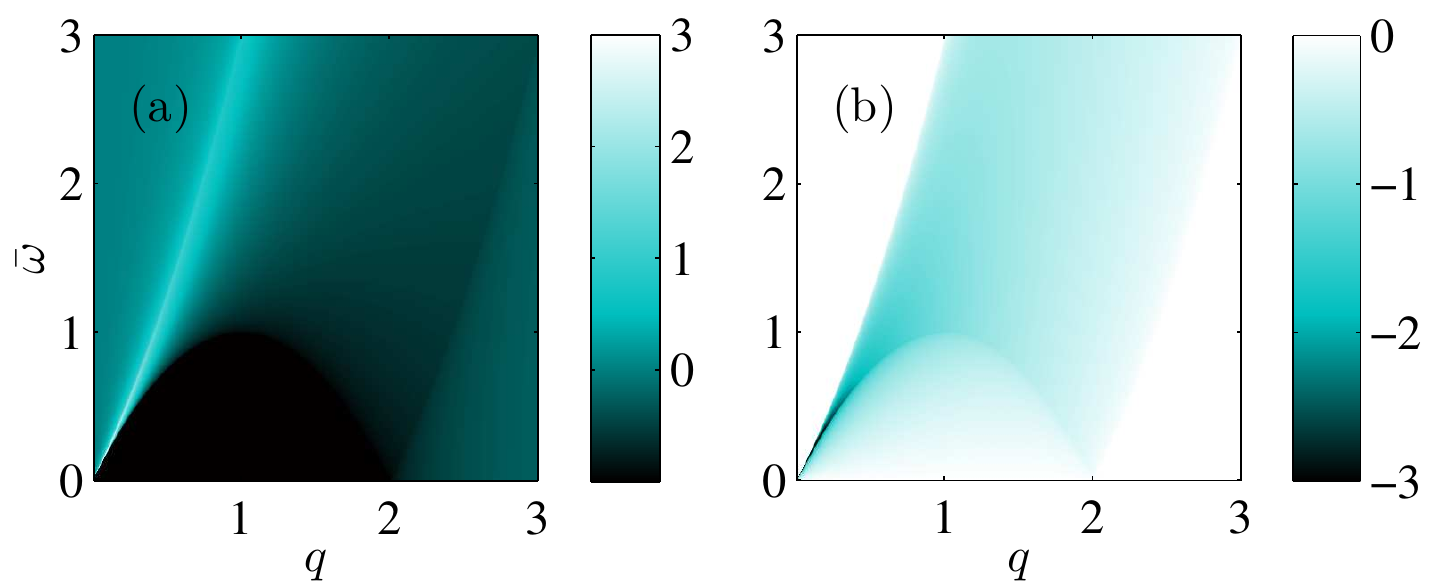}
\caption{Real (a) and imaginary (b) parts of the polarization function for $\bar\mu_n = f_n = g_{2D} = 1$ and $\eta = 0^+$.}
\label{fig:Polarization_Fn}
\end{figure}

By setting $T\rightarrow 0$ and rearranging the terms in Eq.~\eqref{eqn:Pi_Shift}, one gets
\begin{align}
\Pi_n(k,\omega) &=-\int_0^{\sqrt{\mu_n}}dp\oint d\phi \frac{1}{s}\sum_{j = \pm1}\frac{g_{2D}/(2\pi)}{\frac{s^2+j\hbar\hat\omega}{2ps}-\cos\phi}\,.
\end{align}
Here, $g_{2D} = \sqrt{m_xm_y}/(\pi\hbar^2)$ is the two-dimensional density of states. It is convenient to introduce an energy scale $\mu_0$ by defining $\bar\mu_n = \mu_n/\mu_0$, $\bar\omega = \hbar\hat\omega/\mu_0$, and $q = \sqrt{Z}k/\sqrt{\mu_0}$. Performing the angular integral first, followed by the integral along $p$ yields
\begin{equation}
\frac{\Pi_n(q,\bar\omega)}{g_{2D}} = \frac{1}{2}\sum_{j = \pm1}\left[\prod_{l=\pm1}\sqrt{1+\frac{j\bar\omega}{q^2 f_n}+l\frac{2\sqrt{\bar \mu_n}}{q\sqrt{f_n}}}-1\right]\,.
\label{eqn:Pi}
\end{equation}
Setting $\bar\omega\rightarrow0$ recovers the static solution from Ref.~\onlinecite{Low2014pas}. Our dynamic polarization function with $\eta\rightarrow 0^+$ is plotted in Fig.~\ref{fig:Polarization_Fn}.

Using this simplified notation, the interaction term becomes
\begin{equation}
g_{2D}V_{nl}(q) = \sqrt{\frac{2\text{Ha}}{\mu_0}Y}\frac{1}{q}\exp\left(-2Dq\sqrt{X\frac{\mu_0}{2\text{Ha}}}\right)^{|n-l|}\,,
\end{equation}
for $X = m_x/m_e$, $Y = m_y/m_e$, and $D = d/a_0$, where $a_0$ is the Bohr radius.

\section{Collective Modes}

From Eq.~\eqref{eqn:Pi}, one can see that at a given $\bar\omega$, the polarization has an imaginary part for $\sqrt{(\bar\mu_n+\bar\omega)/f_n} - \sqrt{\bar\mu_n/f_n}< q<\sqrt{(\bar\mu_n+\bar\omega)/f_n}+ \sqrt{\bar\mu_n/f_n}$. In other words, the imaginary part vanishes at $\bar\omega>f_n q^2+2\sqrt{\mu_n f_n}q$. This is the region where, according to the RPA formalism, collective modes do not undergo the Landau damping and possess an infinite lifetime.
\begin{figure}
\includegraphics[width = 3in]{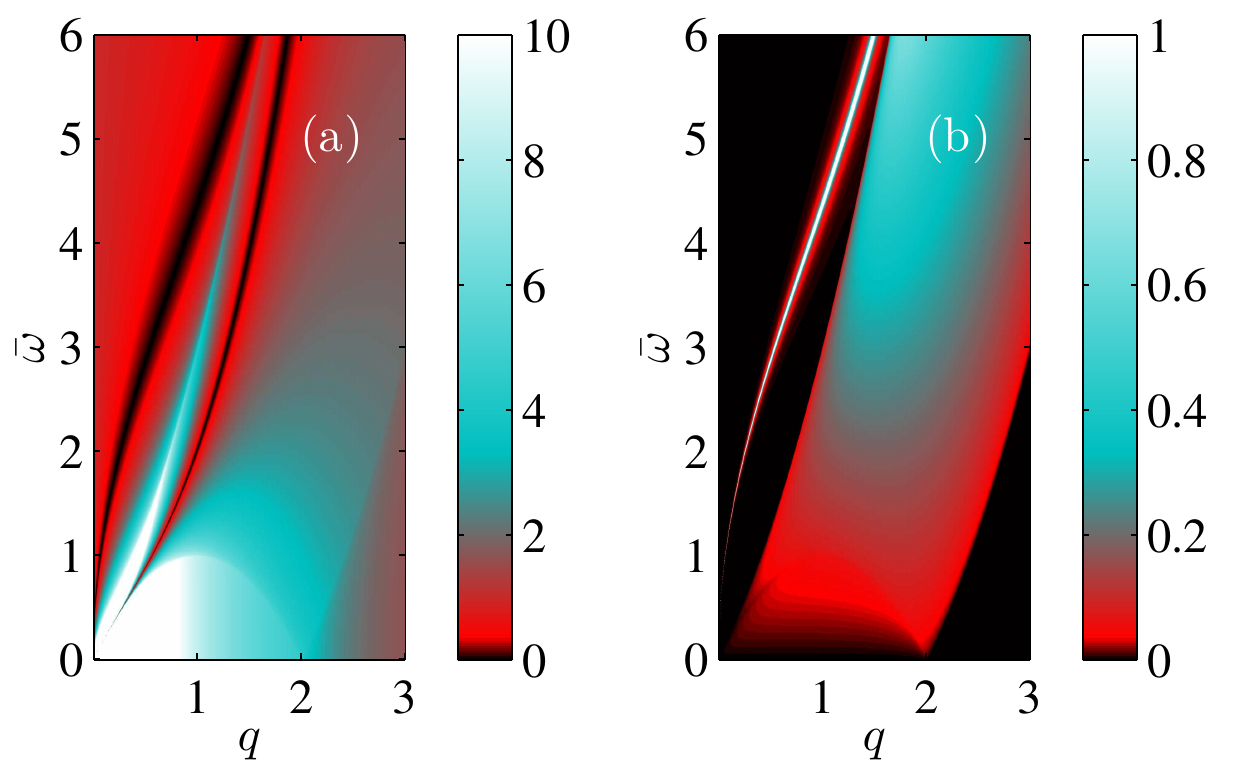}
\caption{$|\Re\text{e}\left[\varepsilon(q,\bar\omega)\right]|$, (a), and $-\Im\text{m}\left[1/\varepsilon(q,\bar\omega)\right]$, (b), for a monolayer system with $\mu_0 = 1$ eV, $\mu_1 = 1$, $Y = 1$, $\theta = \tau = 0$, and $\eta = 10^{-3}$. The plasmonic branch can be seen as the minimum of the absolute value of the real part of the dielectric function and as a peak of the imaginary part of the inverse of $\varepsilon(q,\bar\omega)$. Panel (b) clearly illustrates the particle-hole continuum where the dielectric function has a finite imaginary part.}
\label{fig:Mono_Plasmon}
\end{figure}

Typically, to determine the plasmon dispersion, one solves the determinant equation of the dielectric function numerically. We plot the plasmonic dispersion for this simplest case of a monolayer in Fig.~\ref{fig:Mono_Plasmon}. The general shape of the plasmonic branch agrees with the results from Ref.~\onlinecite{Low2014pas}.

Before we move to the two-layer case, it is instructive to obtain the behavior of the polarization function at small momenta, where $\Pi_n(q,\bar\omega)$ takes a simplified form. For small $ q$, we get
\begin{equation}
\frac{\Pi_n(q,\bar\omega) }{g_{2D}}\approx 2f_n\bar\mu_n\frac{q^2}{\bar\omega^2}\,.
\label{eqn:P_n_Small_q}
\end{equation}
From this, the plasmonic dispersion in a single layer follows a general $\sqrt{q}$ dispersion, which is also seen in graphene monolayers~\cite{Wunsch2006dpo}:
\begin{equation}
\bar\omega=\left( 2f_1\bar\mu_1\sqrt{Y}\sqrt{\frac{\text{2Ha}}{\mu_0}}\right)^{1/2}\sqrt{q}\,.
\label{eqn:Single_Small_q_Plasmon}
\end{equation}

Despite its simple appearence, Eq.~\eqref{eqn:P_n_Small_q} reveals an important feature that is useful for multilayer heterostructures. The rotation/orientation factor $f_n$ acts as a multiplicative modifier of $\bar\mu_n$. To appreciate the importance of this, one should think about the distribution of chemical potentials in stacks of 2D materials. In the presence of a gate, chemical potentials vary between layers monotonically with the layers closest to the gate being impacted the most and screening the more distant sheets. Because of the $f_n$ factor, however, it is possible to have a non-monotonic variation of the \emph{effective} chemical potential $f_n\bar\mu_n$ by rotating the layers with respect to each other. To show how the orientation and relative doping change the low-$q$ dispersion, we solve Eq.~\eqref{eqn:Dielectric_Tensor} for a $2\times2$ matrix and get two plasmonic branches:
\begin{align}
\bar\omega_1&=\sqrt{2\left(f_1\bar\mu_1+f_2\bar\mu_2\right)\sqrt{Y}\sqrt{\frac{2\text{Ha}}{\mu_0}}}\sqrt{q}\,,
\\
\bar\omega_2&=2\sqrt{D\frac{f_1\bar\mu_1f_2\bar\mu_2}{f_1\bar\mu_1+f_2\bar\mu_2}Y\sqrt{M}}q\,,
\label{eqn:omega_2}
\end{align}
where we kept only the leading-$q$ behavior. As expected, one of the solutions shows the standard $\sqrt{q}$ dispersion seen in a monolayer.~\cite{Low2014pas} We will label this branch as SP for ``square-root plasmon". In this case, the geometrically modified chemical potentials $f_n\bar\mu_n$ are added so that the double layer acts as a weighted composite of its monolayer constituents. On the other hand, the second brach has a linear dispersion, referred to as the acoustic plasmon (AP). For this plasmon, the chemical potentials combine to form a reduced chemical potential, similar to a reduced mass. In addition, the existence of this mode depends the separation between the layers $D$ being finite. Similar findings were reported before for graphene~\cite{Hwang2009pmo}; however, the isotropy of graphene dispersion does not permit the geometrical tuning of the plasmonic branches.

\begin{figure}
\includegraphics[width = 3in]{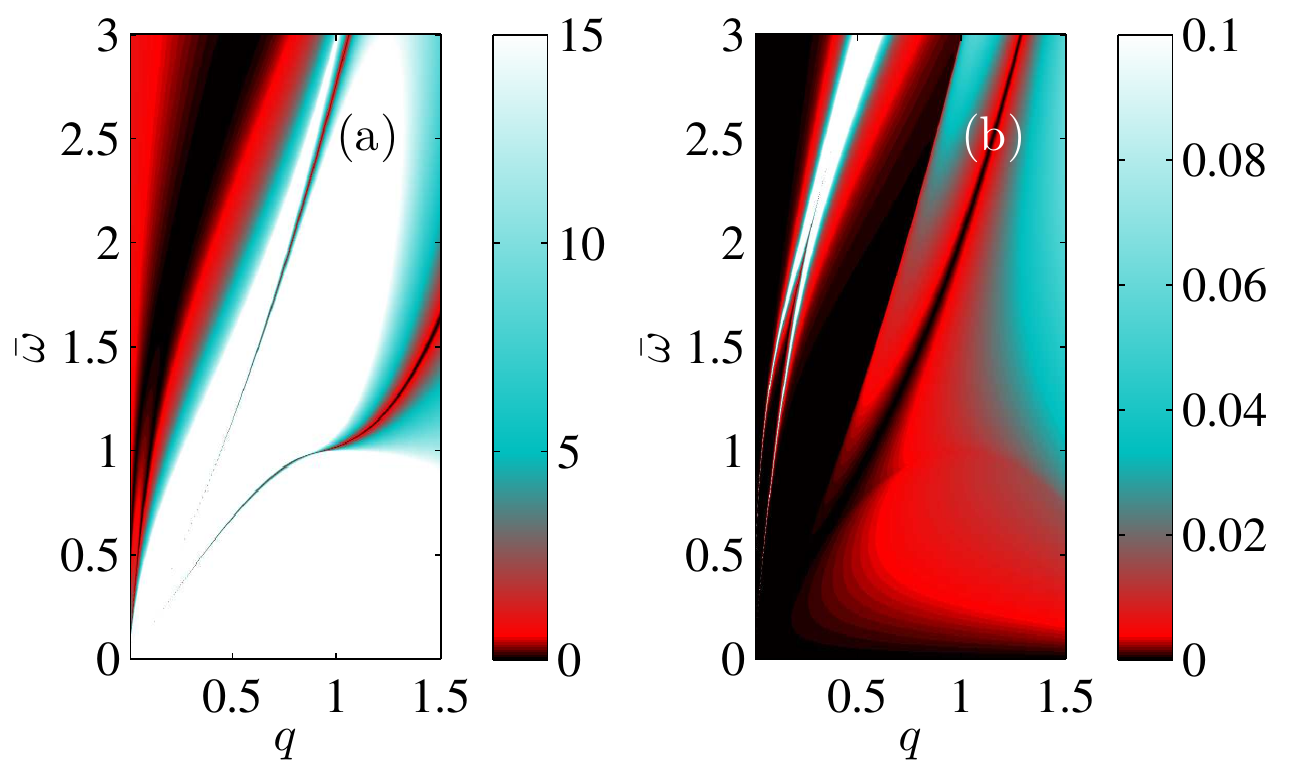}
\caption{$|\Re\text{e}\left[\text{det}(\varepsilon)\right]|$, (a), and $|\Im\text{m}\left[1/\text{det}(\varepsilon)\right]|$, (b), for a rotationally aligned double layer system with $\mu_0 = 1$ eV, $\bar\mu_{1/2} = 1$, $Y = 1$, $\theta =  0$, $D = 100$, and $\eta = 10^{-3}$.}
\label{fig:Double_Plasmon}
\end{figure}

To get a fuller picture, we plot the determinant of the dielectric tenson in Fig.~\ref{fig:Double_Plasmon}. One can see that the branches are separated at low momenta, but become indistinguishable at higher values of $q$. The reason for this, in addition to the finite broadening, is the fact that the coupling between the layers decays exponentially with $q$. This means that at large $q$, the layers become essentially independent and since they are identical, the branches merge. At high-enough momenta, the plasmons enter the particle-hole continuum and acquire a finite lifetime due to the Landau damping.

\begin{figure}
\includegraphics[width = 3.4in]{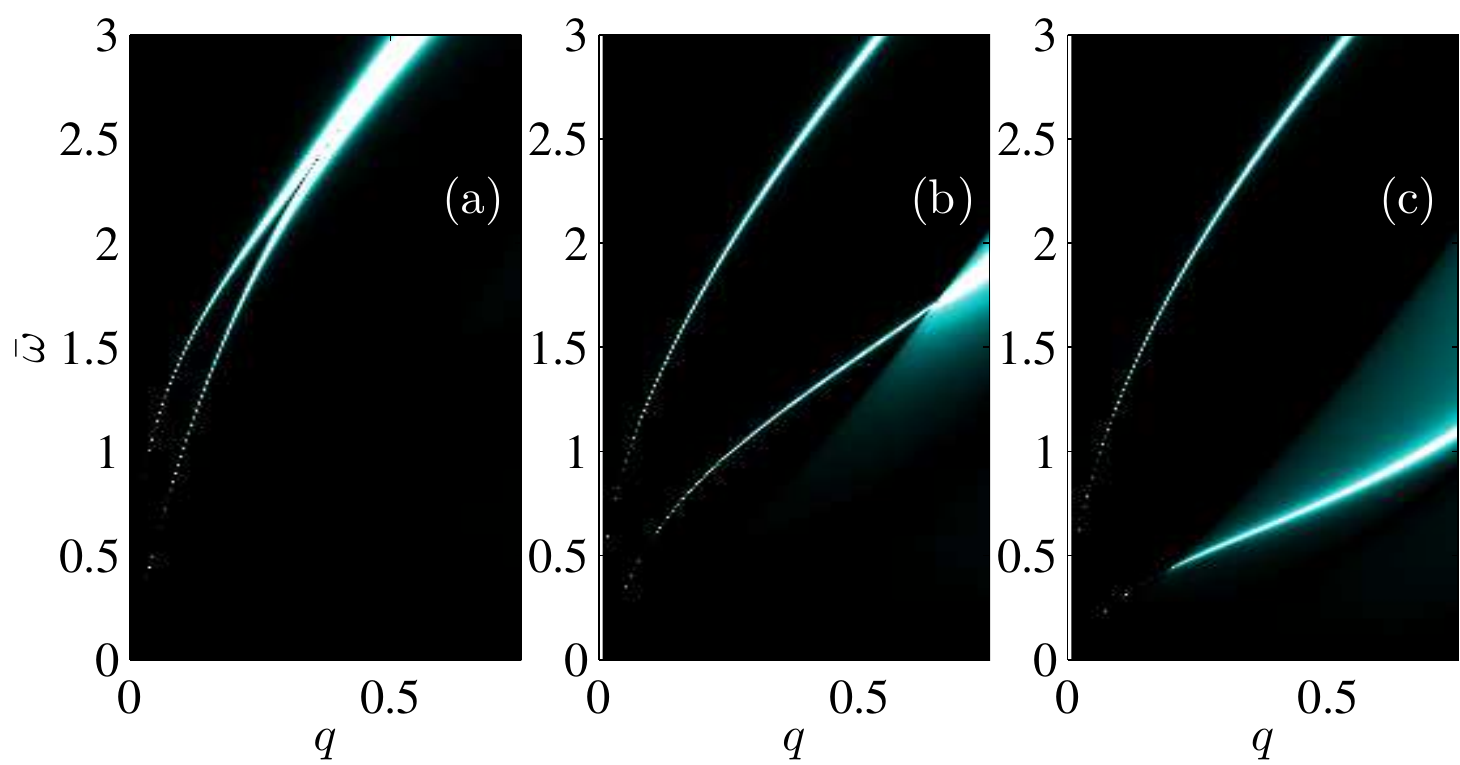}
\caption{$|\Im\text{m}\left[1/\text{det}(\varepsilon)\right]|$ for the aligned double layer system with $\tau_{1/2} = 0$, $\mu_0 = 1$ eV, $Y = 1$, $M = 1/8$, $D = 100$, $\theta = 0$, $\bar\mu_1 = 1$, and $\bar\mu_2 = 1$ (a), 1/2 (b), 1/4 (c).}
\label{fig:mu_Var}
\end{figure}

In order to understand how the numerous parameters impact the behavior in the double layer system, it is helpful to see the variation of the plasmonic branches as we tune the variables. We start with the simplest case of rotationally aligned layers with unequal doping, Fig.~\ref{fig:mu_Var}.
We show three situations in which $\bar\mu_1 = 1$ and $\bar\mu_2$ is progressively lowered from 1 to 1/2 to 1/4. As before, there are two plasmonic branches. It is immediately clear that one of the branches is much more sensitive to the variations in the chemical potential. While the SP undergoes only a very slight modification, the AP changes significanly with the reduction of $\bar\mu_2$.   The AP is more gradual at smaller $\bar\mu_2$ and merges with the particle-hole continuum at lower $q$. This agrees with Eq.~\eqref{eqn:omega_2}, where $\bar\omega_2\rightarrow0$ for a vanishing $\bar\mu_2$.

\begin{figure}
\includegraphics[width = 3.4in]{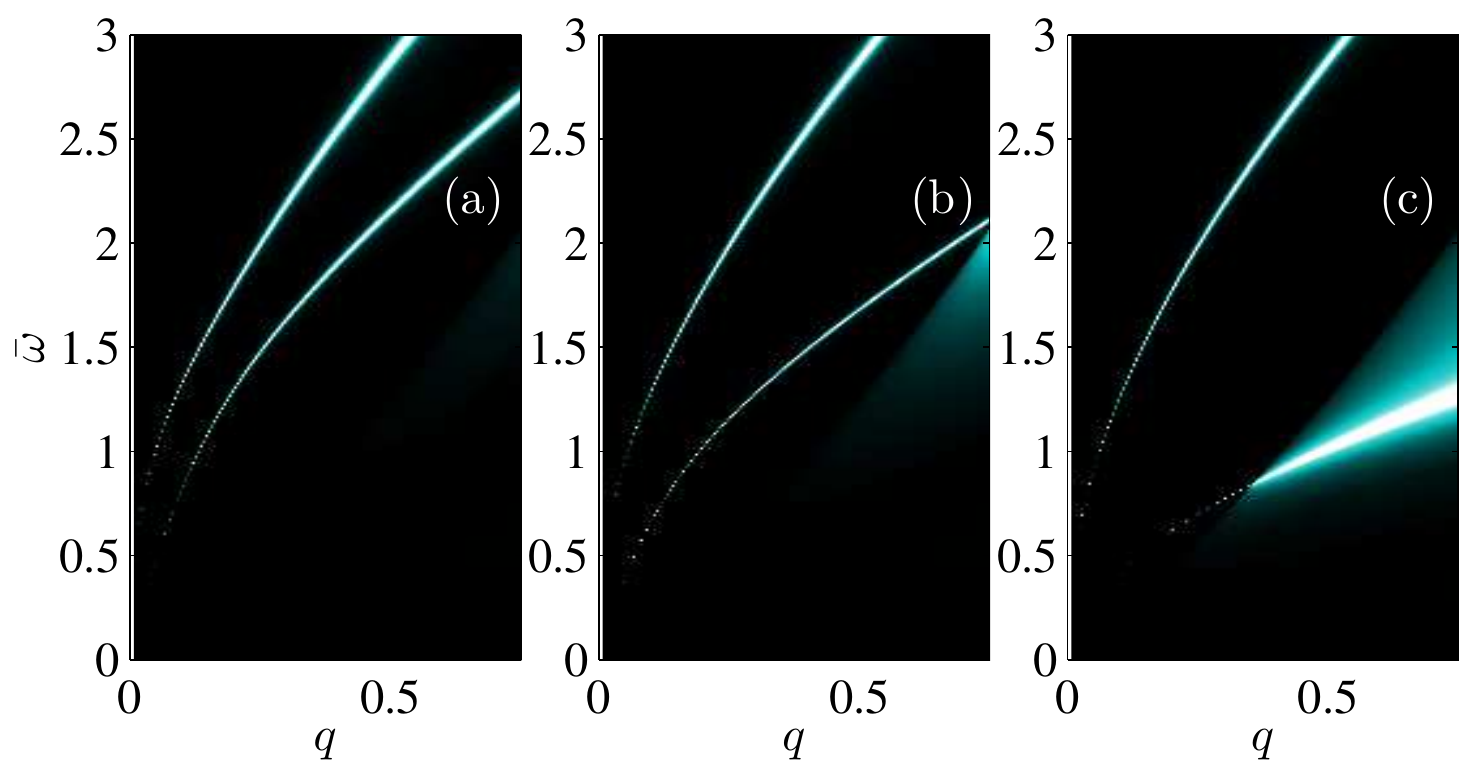}
\caption{$|\Im\text{m}\left[1/\text{det}(\varepsilon)\right]|$ for a double layer system with $\tau_1 = 0$, $\mu_0 = 1$ eV, $Y = 1$, $M = 1/8$, $D = 100$, $\theta = 0$, $\bar\mu_{1/2} = 1$, and $\tau_2 = \pi/4$ (a), $\pi/3$ (b), $\pi/2$ (c).}
\label{fig:tau_Var}
\end{figure}

Next, we address the effect that rotation has on the system where $\bar\mu_1 = \bar\mu_2$, see Fig.~\ref{fig:tau_Var}. Here, increasing the twist angle has a similar effect as the reduction of $\bar\mu_2$ in the previous case. The reason is that for $M<1$, which is what we use, greater twist leads to a smaller value of $f_2$. According to Eq.~\eqref{eqn:omega_2}, this is indeed equivalent to reducing $\bar\mu_2$. From this, it is clear that the geometric factor $f_n$ determines the system's sensitivity to the changes in the chemical potential. Just like for varying $\bar\mu_2$, only the AP experiences a substantial modification at different angles of rotation with SP remaining fairly unchanged.

\begin{figure}
\includegraphics[width = 3.4in]{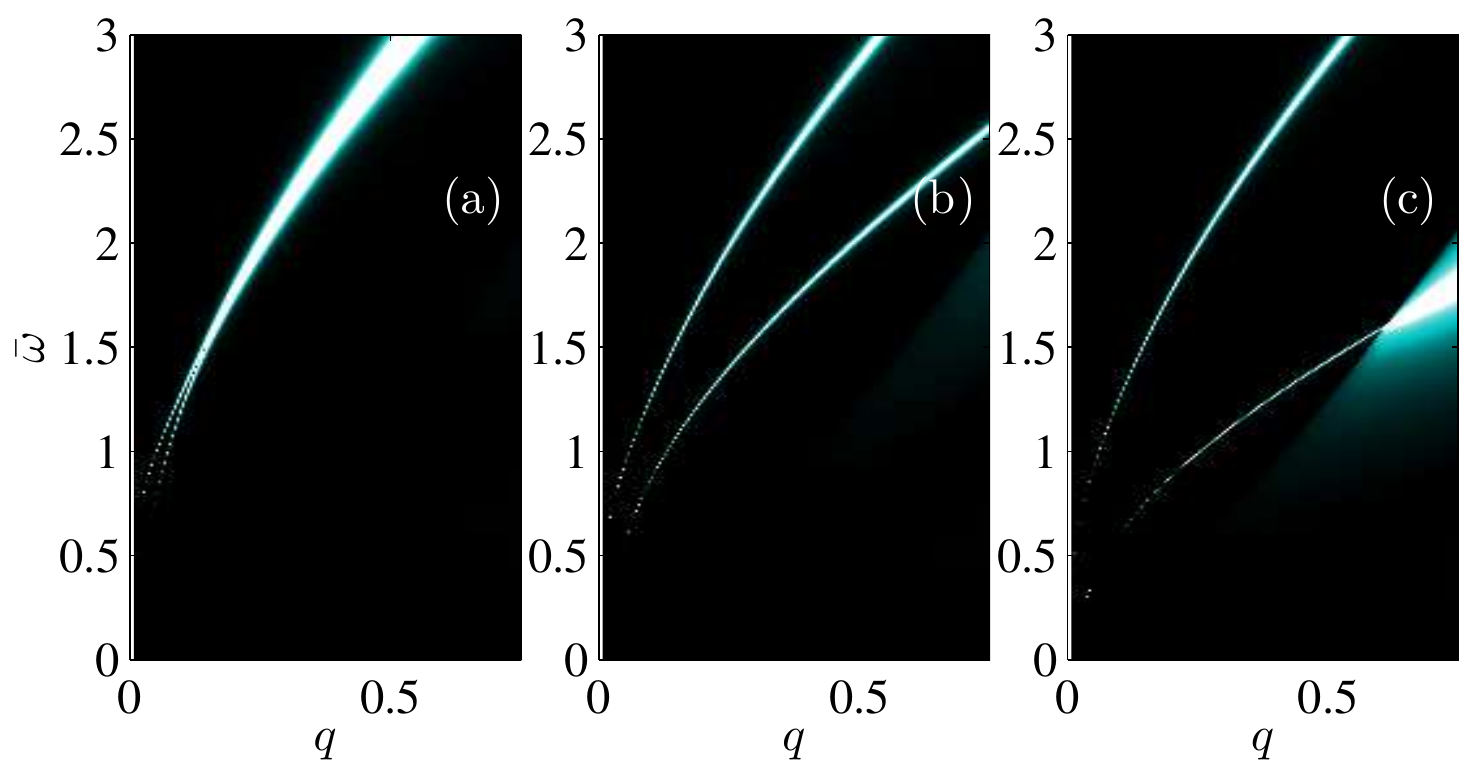}
\caption{$|\Im\text{m}\left[1/\text{det}(\varepsilon)\right]|$ for a perpendicular double layer system with $\tau_1 = 0$, $\tau_2 = \pi/2$, $\mu_0 = 1$ eV, $Y = 1$, $D = 100$, $\theta = 0$, $\bar\mu_{1/2} = 1$, $M = 1$(a), 1/2 (b), 1/4 (c).}
\label{fig:M_Var}
\end{figure}

The final parameter that we consider is the anisotropy factor $M$. We plot the plasmonic dispersions for equally doped perpendicularly oriented layers for several values of $M$ in Fig.~\ref{fig:M_Var}. As expected, larger anisotropy leads to a greater variation in $f_n$ as a function of direction. 

From Figs.~\ref{fig:mu_Var}--\ref{fig:M_Var} one can see the following sensitivity hierarchy. Anisotropy measure $M$ determines the geometric factor's, $f_n$, sensitivity to misalignment. The factor $f_n$, in turn, controls the dependence of the dispersion on relative doping in the two layers. Naturally, the system most susceptible to variation in doping would be highly anisotropic and perpendicularly oriented.

\section{Conclusions}
To summarize, we have derived a dynamic polarization function for a massive anisotropic system. From this, it is possible to obtain collective modes in coupled, but isolated, multilayer systems with arbitrary orientation and doping. Here, we focused on the simplest case of a double layer system.

By finding the zeros of the dielectric tensor, we obtained the plasmonic dispersion for a range of parameters. In particular, we showed the dependence of the plasmonic modes on the system anisotropy, relative orientation, and relative doping. Our most important finding here is that the anisotropy introduces a new control knob which can be used to obtain the desired behavior in multilayer systems. Given that there already exists a layered anisotropic semiconductor, we believe that our results can be of a direct benefit to the experimental community.

 A.H.C.N. and A. S. R. acknowledge the National Research Foundation, Prime Minister Office, Singapore, under its Medium Sized Centre  Programme and CRP award ”Novel 2D materials with  tailored properties: beyond graphene” (R-144-000-295- 281).

\end{document}